# Asymmetric multi-segmented conjugated polymer–metal nanowires for engineering of non-linear electrical behavior

Loïk Gence, Vincent Callegari Sophie Demoustier-Champagne and Jean-Luc Duvail*

Institut de la Matière Condensée et des Nanosciences - Bio & Soft Matter

(IMCN/BSMA), Université catholique de Louvain, Place Croix du Sud, 1, B-1348 Louvain-La-Neuve, Belgium

*Institut des Matériaux Jean Rouxel, CNRS, Université de Nantes, Nantes, France

loik.gence@cetuc.puc-rio.br

ABSTRACT  In this paper, we report on the impact of metal/polymer interfaces on the rectifying behavior of novel multi-segmented hybrid nanowires (HNWs) made of metallic and conjugated polymer (CP) segments. Using HNWs integrated in micromachined devices, the relationship between electronic properties and original structure is revealed. By combining transmission electron microscopy (TEM) and current-voltage (*I-V*) spectroscopy studies performed on several symmetric and asymmetric HNWs structures, we show that rectifying *I–V* characteristics are observed only for asymmetric HNWs. Moreover, it is shown that the rectification ratio can be improved up to 3 orders of magnitude by a proper selection of the HNW composition. While the rectifying behavior is observed in HNWs after oxidative or acid doping, the charge transport mechanism in as-synthesized HNWs is bulk-limited and independent from their structure. Both symmetric and asymmetric HNWs exhibit Ohmic and non-linear *I-V* curves above and below $T_{NL}$ ≈120 K, respectively. These electrical behaviors are consistent with a smooth transition from an Ohmic to a non-Ohmic variable-range-hopping (VRH) mechanism. We discuss the origin of these nonlinearities comparing the two- and four-probe measurements on single HNWs and we propose a simple model based on dual back-to-back Schottky diodes to explain qualitatively the rectifying properties.

## 1. Introduction

In the last few decades, —many studies have been devoted to the synthesis and the characterization of conjugated polymers (CPs) nanowires (NWs) and nanotubes (NTs) because they combine (opto)electronic properties(see for reviews [1-3,4]) of inorganic semiconductors and metals together with the good processability of polymers, which confers on them an impressive potential for applications, ranging from bio-environmental sensing [2,5] to high density memories and supercapacitors [5]. CPs are also envisaged in flexible electronics, for answering the growing demand for new electronic devices featuring low fabrication costs and compatibility with transparent and flexible substrates [6]. One of the key challenges is to develop novel memory technologies that fit these requirements. In that respect, CPs have been proposed as active materials for resistive switching memories (RSM) [7]. In a typical RSM structure, the active material is sandwiched between two metallic electrodes. While organic RSMs must still be improved in terms of performances, especially when compared to inorganic RSMs, they can take advantage of the unique features of CPs.

Intriguingly, while CPs present vast technological opportunities and can be easily combined with inorganic materials, hybrid metal-CP NWs have scarcely been studied as most of the studies focused on purely inorganic multi-segmented NWs [8]. Remarkably, metal-CP hybrid nanowires (HNWs) can easily fit the device structure found in typical two-terminal CP resistive memory devices [9]. Among HNWs, only symmetric tri-segmented metal-CP-metal structures have been investigated [10-13], although more complex HNWs can be produced by high-throughput template strategies which offer a very good control over the composition and spatial distribution of the different segments. We believe that HNWs could be used to develop nanoscale CP resistive memory devices.

Recently, we detailed the advances in the preparation of a variety of multi-segmented nanowires containing a combination of metallic and polymeric components [14,15]. We showed how to overcome some of the difficulties, in particular for building strong nanoscale interfaces between the different segments. To achieve the integration of CP NWs and HNWs in functional devices, it is still necessary to properly control their electrical features. A better understanding and control of the charge transport mechanisms involved in phenomena like resistive switching is also required [6,7].

These 1D-like structures are of great interest for investigating transport mechanisms at the nanoscale, where the quantum confinement effect, the charge tunneling, the Coulomb charging energy, the electron-electron interactions (EEI), and the disorder all play important roles [1,16-17]. The interest in CP NWs exploded after the seminal work of Martin et al [18] who highlighted for the first time the improvement of molecular and supramolecular structures of CP films made of template-grown CP compared to films prepared by conventional methods. Since then, the transport properties of NWs and NTs have been investigated for various CPs such as polypyrrole (PPy), poly(3,4-ethylenedioxythiophene) (PEDOT) polyacetylene (PA) or polyaniline (PANi) as a function of parameters such as temperature [19-21], type of dopant [22], magnetic field [23,24] or electrical field intensity [19, 25], to give only a few examples. However charge transport in CP NWs still remains a topic of significant controversy because of the strong influence of the synthesis techniques and conditions on the transport properties of these materials. As a result, various theoretical models have been used to describe charge transport in CPs (see Ref. [1,4, 16,19,20, 27] and references therein).

In this paper, we focus on the transport properties of tri-, tetra- and penta-segmented HNWs recorded in both planar and vertical configuration. We show that among HNW structures, asymmetric HNWs present the interesting property to switch their *I-V* characteristics from a linear to a rectifying behavior upon chemical (acid and oxidative) secondary doping. Interestingly, the rectifying ratio can be improved (up to a factor of ~3600) by a proper selection of the HNW composition. The origin of these nonlinearities are discussed, based on the chemical analysis and by comparing the two- and four-probe measurements of single HNWs. We propose a simple model to explain qualitatively this behavior. To complete this study, we closely scrutinize and compare the transport properties of the as-synthesized symmetric and asymmetric HNWs over a wide range of temperature $T$, electric $F$ and magnetic $B$ fields. We show that they are independent from their (a)symmetry and composition. All these results suggest that the observed nonlinearities are induced by a smooth transition from an activated variable-range hopping (VRH) mechanism to a $F$-induced hopping mechanism dominating in the low-$T$ regime ($T \leq 30$ K).

## 2. Results and Discussion

Figure 1 gives schematic views of the symmetric (a) and asymmetric (b) multi-segmented HNWs studied in this work. The electrical characterization of such template-synthesized NWs can be performed in a vertical configuration within the template, or in planar configuration with NWs dispersed on a substrate, as exemplified in figure 2-(a) and (b). The vertical configuration is quite convenient and it has been frequently employed for the electrical [17,28,29,30-32] and thermal [32a,33] characterizations of various types of NWs as it requires no complex post-synthesis processing of the sample to be measured. However, the electrical probing of a single NW and its response for example to chemical stimulus can only be achieved after the template removal, *i.e.* by the planar configuration. For contacting single HNWs in the planar configuration, oxidized silicon wafer and membrane-based micromachined devices were used as substrates. Thanks to a transparent silicon nitride window, the microdevices (Fig. 2.c) enable the correlation of the electrical data to the structural and the chemical analyses obtained from SEM, TEM or AFM characterization. This allows a straightforward understanding of the structure-function relationship of our samples. Elemental composition of each CP segment can be obtained by energy-dispersive X-ray (EDX) analysis to unambiguously identify the polymer segments: polypyrrole (PPy) or poly(3,4-ethylene dioxythiophene) (PEDOT).

Figure 2 shows an example of EDX spectra recorded on the three spots marked in Figure 2b. These spectra reveal the presence of sulfur atoms, characteristic for PEDOT, only in the spot III. Both bottom-contact and top-contact methods were used for depositing the electrical leads. Each method has its own advantages and disadvantages [4]. The main advantage of the top-contact approach using EBL is to allow the exact positioning of multiple nanoelectrodes on top of the HNWs. Thus, it allows the measurement of small portion of HNWs, which is crucial for nonuniform nanostructure like the HNWs shown in Fig. 1. HNWs were synthesized by sequential deposition of three, four and five metallic and CP (PPy, PEDOT, PANi) segments within the pores of 20 µm thick polycarbonate (PC) membranes - with pore diameter ranging from 90 to 110 nm and a pore density of $10^9/cm^2$, using an all-electrochemical process [14]. In a two-probe configuration, gold microelectrodes are deposited on the two extreme metal segments while other electrodes are also deposited on the inner segments for the four-probe configuration. The comparaison of the two- and four-probe measurements allow determining the impact of the synthesized metal/CP interfaces on the $I$-$V$ curves of single HNWs. It is worth of noting that this EBL processing might lower the doping level of CP segments [34]. The chemical doping of HNWs, —also called secondary doping, was performed by immersing single NW integrated into devices in an aqueous solution containing $FeCl_3$ 0.5 M or a HCl 1 M solution for 1 hour. Typical $I$-$V$ characteristics of asymmetric and symmetric HNWs at room temperature (R$T$) are given in Fig. 3 (a) and (b), respectively. The dashed and solid curves correspond respectively to the as-synthesized and doped states. The figure 3-(a) shows the two- and four-contacts $I$-$V$ characteristics of a single Au-PEDOT-Au HNW before and after the oxidative treatment. The resistance, estimated from the slope of the *I-V* curves around zero-bias point, is lowered after doping roughly by a factor of 1.4 and 2 for the two- and four-contact measurements, respectively. This resistance decrease is usual for p-type CPs such as PPy and PEDOT, and it stems from the lowering of the intrinsic resistance of the bulk (distant from the interfaces) of CP segments, denoted $R_{PPy}$ and $R_{PEDOT}$ in the scheme of the Fig. 3 (e). This is a consequence of the transfer of charges to the CP chains after oxidation. It is worth of note that in a two-probe measurement, the overall conductivity increase of HNWs after doping results from two competing effects. On one hand, the bulk resistance of the p-type CP is expected to decrease upon oxidative treatment. On the other hand, the metal/CP interface resistance may increase because of possible shape and area modifications of the metal/CP interfaces or the appearance of small Schottky barriers. These additional interface resistances explain the difference between the two- and four-probe measurements. It is worth of note that most of our samples exhibited a resistance decrease after as shown in Fig.3.(a-b). However, a small fraction of HNWs experienced an overall resistance increase because of the unavoidable dispersion of the CP doping level and interface quality. Besides the modification of the conductivity of the CP segment, a change from Ohmic to rectifying *I-V* characteristics is observed for asymmetric HNWs as shown in Fig.3(b). The rectifying ratio (*RR*), defined as the ratio of forward to reverse currents, is varying between 2 and 3 for au-PEDOT-PPy-Au HNWs. This effect of the structural asymmetry is clearly highlighted when comparing the *I-V* curves recorded for single Au-PEDOT-Au and Au-PEDOT-PPy-Au HNWs shown in the panels a and b of Fig. 3.

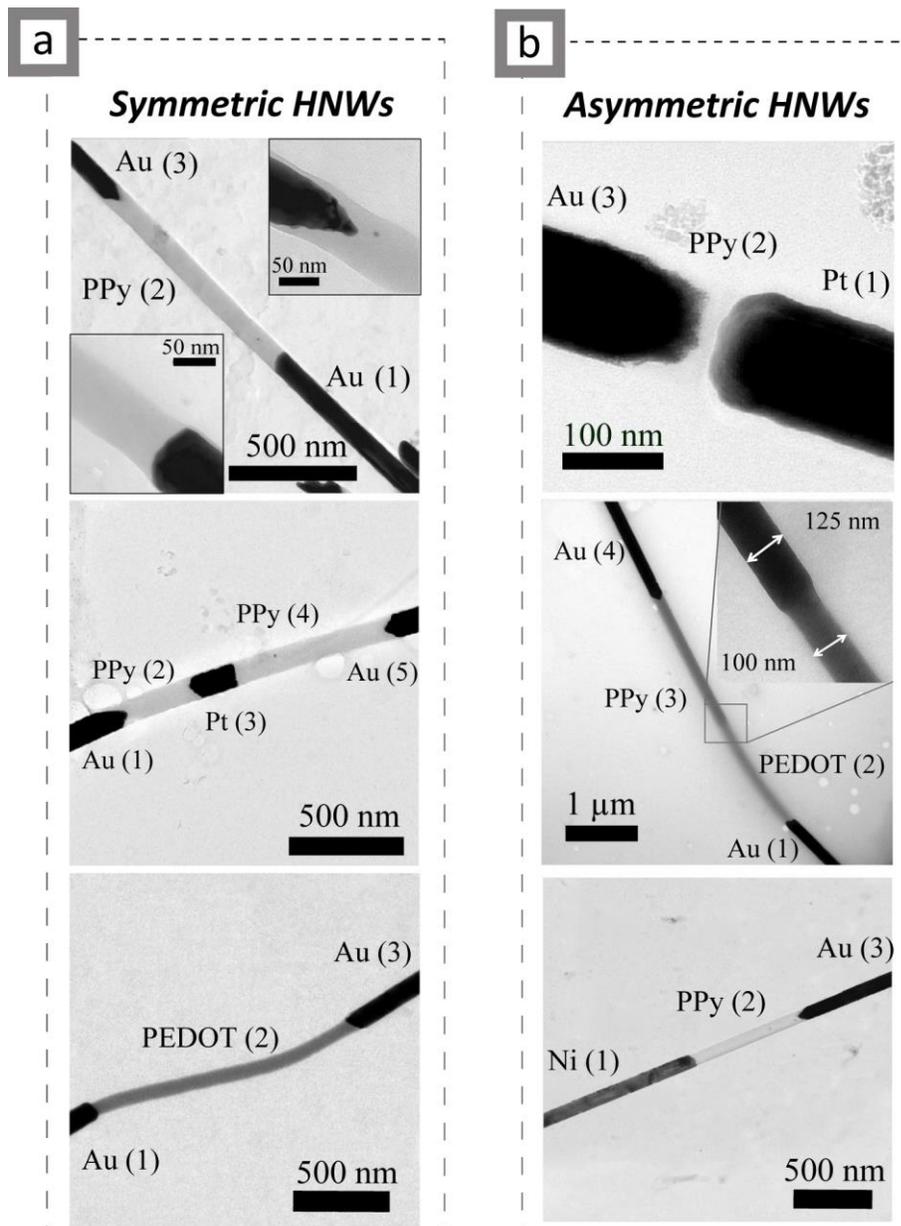

**Figure 1 Transmission electron microscopy (TEM) images of symmetric (a) and asymmetric (b) HNWs synthesized by an all-electrochemical method and studied in this work. The numbers indicate the synthesis order.**

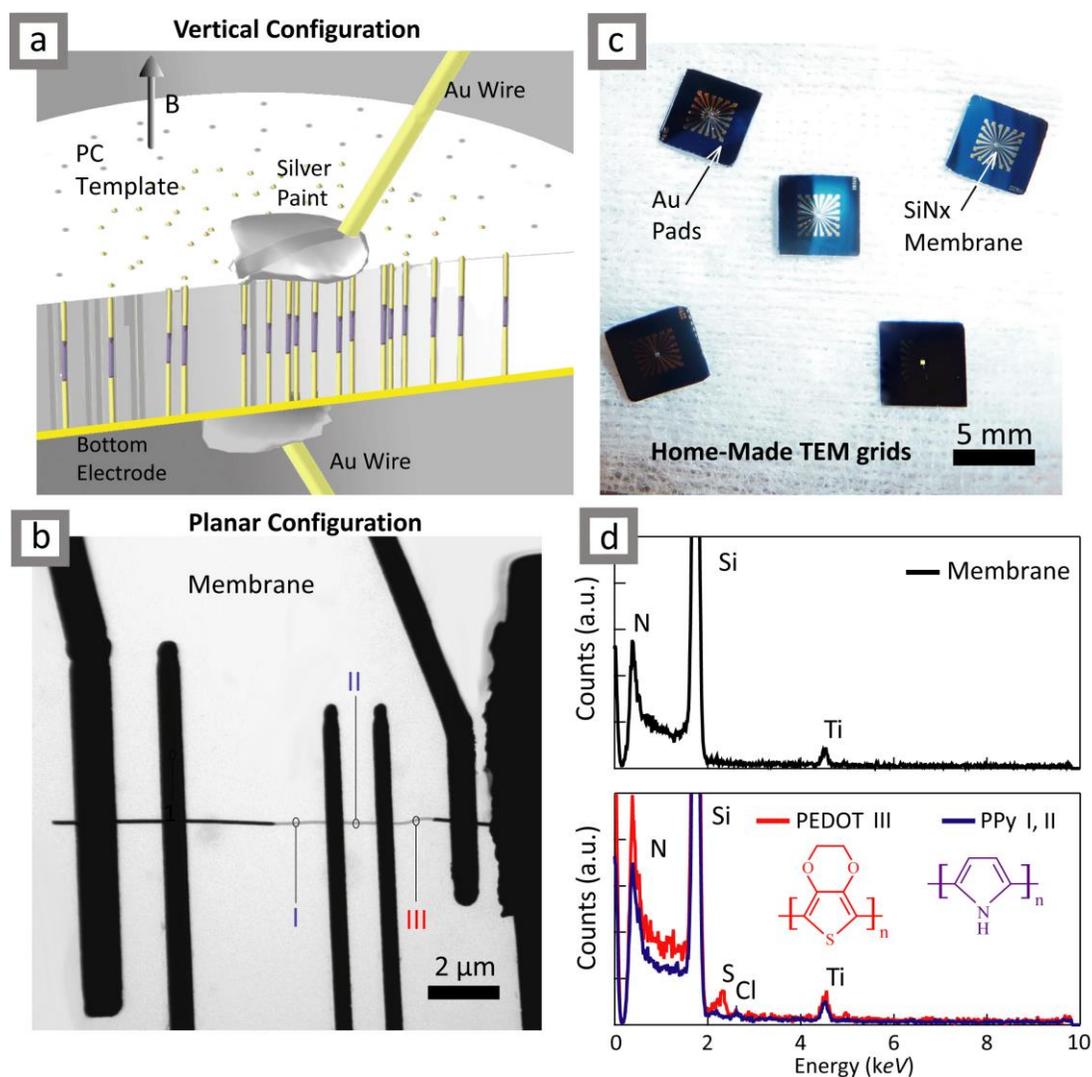

**Figure 2** HNWs were characterized in a vertical (a) or planar (b) configuration using membrane-based microdevices (c). To unambiguously distinguish the polymer segments, we employed energy-dispersive X-ray (EDX) analysis. The EDX spectra recorded on the three spots reveal the presence of sulfur atoms, characteristic for PEDOT, only in the area III.

In a two-point configuration, the measured resistance is the sum of the intrinsic resistance of the segment comprised between the two probes, and the contact resistance composed of the lead resistances and the lead/HNW interface resistances [35]. The rectifying effect is recorded only in two-point measurements. One can reasonably assume a role of the metal/CP interfaces. The importance of metal/CP interfaces in the measurement of HNWs was already highlighted in a previous paper where we showed that a single monolayer of SH−(CH$_2$)$_3$−COOH grown onto the first Au segment modifies drastically (by almost 3 orders of magnitude) the zero-bias resistance of tri-segmented HNWs [36]. Strikingly, Au-PEDOT-Au HNWs always exhibit symmetrical *I-V* curves, before and after doping, independently from the probe configuration (i.e. two or four probes). The rectifying behavior is thus related to the difference in the HNW structure. To investigate further the role of metal/CP interfaces, we probed different segments of as-synthesized Au-PEDOT-PPy-Au HNWs in a four-point configuration, as shown in Fig 3(c) where the inset gives TEM pictures of both metal/CP interfaces. It is noteworthy that the interfaces exhibit very distinct shapes and areas induced by the sequential deposition of the HNW segments. Panel 3(d) shows that the *I-V* curves follow a linear relationship for PEDOT/PPy (case 1), Au/PEDOT (case 2) and PPy/Au (case 3) when as-synthesized. The data suggest that as-synthesized Au/PEDOT and PPy/Au interfaces are not equivalent in terms of electrical conductivities. The Au/PPy interface (case 3) presents a conductance roughly 15 times smaller than the Au/PEDOT (2) interface and 24 times smaller than the CD segment (1).

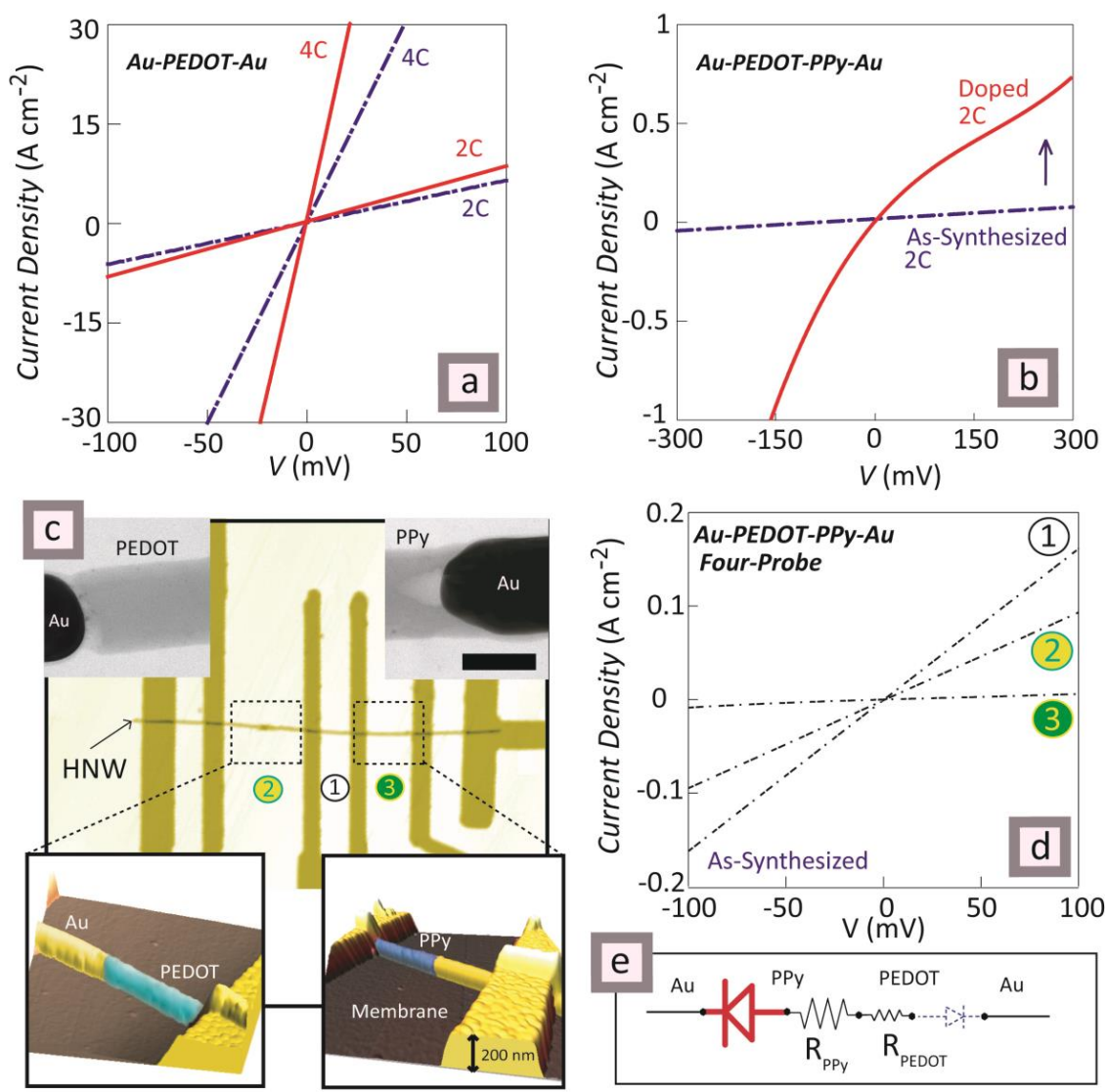

**Figure 3 (a-b)** *I-V* characteristics measured for as-synthesized (dashed curve) and doped (solid curve) (a) asymmetric and (b) symmetric HNWs at room temperature. (c) AFM image of an asymmetric HNW contacted by several electrodes on a home-made TEM grids. The lower insets gives a 3D view of the metal-CP interfaces while the upper insets give the corresponding TEM micrographs. The scale bar for TEM images is 100 nm. (d) four-probe measurements recorded for each segment identified in (c) in the as-synthesized state. (e) Equivalent electrical circuit of a doped asymmetric Au-PEDOT-PPy-Au HNW.

Besides different metal/CP contact areas, the existence of small Schottky barriers can also be invoked to account for the variation in the resistance of the metal/CP interfaces. Highly doped P-type CPs like PPy and PEDOT with a work function (WF) $\Phi_{CP}$ are expected to form Ohmic contacts with metals featuring a higher WF ($\Phi_M$) such as Au and Pt. In the opposite case, even a small WF difference causes the emergence of a Schottky barrier ($\Phi_B$) at the metal/CP interfaces. However, we already showed in a previous work [36] that charge transport in HNWs is not electrode-limited even at low temperatures (T<100 K) confirming that Schottky barriers are negligible ($\Phi_B$ <10 meV) in as-synthesized Au-PPy-Au HNWs. Schottky barriers were often used in the description of two-terminal NW devices [37,38], especially in junction-based bio- and gas sensor applications [38-41]. Usually, the charge injection at metal/CP interfaces featuring a Schottky barrier $\Phi_B$ is described by the thermoionic emission theory [42] and the current density $J$ is exponentially related to the work function of the CP

$$J = A^* T^2 \exp(\frac{-\phi_B}{k_B T}) \exp(\frac{eV}{n k_B T}) \qquad (1)$$

As a result, each interface presents a zero-voltage specific contact resistivity of the form [43]

$$\rho^{M-CP} = \frac{k_B}{A^* T q} \exp(\frac{q \phi_B}{k_B T}) \qquad (2)$$

where $A^* = 4\pi q m^* k_B^2 / h^3$ is the effective Richardson constant with $m^*$ the carrier effective mass, $T$ the absolute temperature, $k_B$ the Boltzmann constant and n the ideality factor. Because of the exponential dependence of $\rho^{CP}$, even a low Schottky barrier at the PPy/Au interface give rise to a consequent contact resistance shift. Thus, despite the apparent Ohmic $I$-$V$ curve, the existence of a Schottky barrier at the PPy/Au junction could be responsible for its high resistance. This is well corroborated by the measured conductivity distribution for Au/PEDOT/Au and Au/PEDOT/PPy/Au HNWs (See SI). It is clear that this asymmetry of the as-synthesized PPy/Au and PEDOT/Au junctions plays a key role in the rectifying behavior observed for tetrasegmented HNWs. The lower inset of Fig.3.(e) is a schematic diagram of the equivalent circuit of a Au-PEDOT-PPy-Au HNW with two asymmetric Schottky barriers. For such metal-semiconductor-metal (MSM) structure, the current density is limited for each bias polarity by the corresponding reversed-biased Schottky contact. From equation 1, it is immediate to see that the rectifying ratio (*RR*) for a MSM structure like the one of Fig. 3(e) depends exponentially on the difference in barrier heights $\Delta \phi = (\phi_B^{CP1} - \phi_B^{CP2})$.

However $\Delta\Phi$ is very difficult to estimate as the effective barrier heights at both metal/organic interfaces are strongly influenced by the presence of interfacial dipoles between the metal surface and organic layer [43,44] and by the exact work function of the metal which greatly depends on its surface properties as demonstrated by the wide range of values found in the literature for Au and Pt. In addition, conduction mechanisms other than thermoionic emission are expected to occur in nano-interfaces depending on the shape and width of the barrier [45,46]. The Ohmic behavior observed for as-synthesized HNWs demonstrates that Schottky barrier heights and $\Delta\Phi$ are small enough to be neglected in as-synthesized HNWs. On the other hand, the barrier heights and asymmetry are strongly enhanced after the oxidative treatment. This is well supported by the XPS analysis showing that PPy and PEDOT undergo very different doping level and chemical alterations after $FeCl_3$ treatment [15]. We showed that PPy is greatly affected by $FeCl_3$ exposure and exhibit a strong modification of its doping ratio and electronic structure after $FeCl_3$ treatment while PEDOT is almost not affected. These modifications impact the PPy and PEDOT work functions which reflect on the corresponding Schottky barriers at the PPy/Au and PEDOT/Au junctions.

To deepen our understanding of the rectifying behavior in asymmetric HNWs, we synthesized and contacted single asymmetric HNWs containing a Pt/PANi and a PPy/Au junction. Because the conductivity of PANi can be easily modified (typically more than 6 orders of magnitude [47]), PANi associated to PPy is a good choice to strengthening the rectifying effect in HNWs. Aiming to improve *RR*, the oxidative treatment was replaced by protonation (HCl 1M), as it is known to be an excellent means of increasing the conductivity of PANi while on the other hand, PPy conductivity is expected to be only slightly modified by such protonation [47]. The two-probe *I-V* curves of single Pt-PANi-PPy-Au HNWs with electrodes on Au segments, before and after protonation, are given in Fig. 4(c). Similarly to pristine Au-PEDOT-PPy-Au HNWs, symmetrical *I-V* characteristics are observed before protonation while a strong rectifying behavior is observed after protonation. For each bias polarity, one metal/CP junction is forward biased and the other one is in reverse mode. Remarkably, as-synthesized HNWs exhibit Ohmic

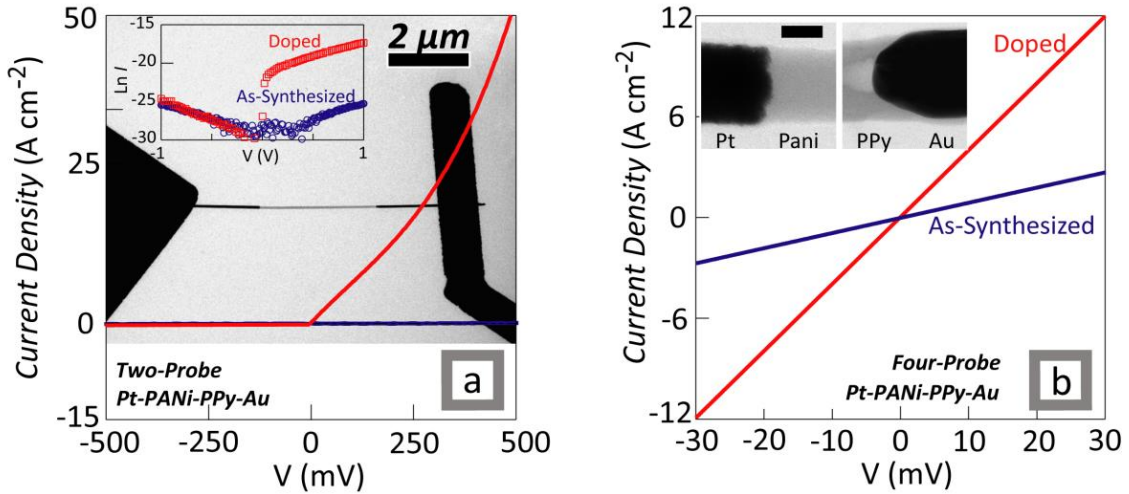

**Figure 4** *I-V* **curves for Pt-PANi-PPy-Au HNWs recorded before and after protonation, in a two-probe configuration. The inset shows the data in logarithmic scale. (d) Four-point measurements of the CP segment (voltage probes on PPy and PANi segments). The scale bar represents 50 nm.**

*I-V* characteristics (*RR*=1), meaning that the barrier heights $\phi_B^{PPy}$ and $\phi_B^{PANi}$ are small and their difference negligible $\Delta\phi \approx 0$. On the other hand, because of the strong response of PANi to protonation, its effective barrier height is cancelled and the usual *I-V* characteristics of a Schotkky diode is obtained after protonation as shown in Fig. 3-(c). At 150 mV, a rectification ratio *RR* = 4 was found with Au- PEDOT-PPy-Au HNWs, whereas a rectification ratio 3 orders of magnitude higher (RR=3600) was observed for the Pt-PANi-PPy-Au structure.

While these values are recorded directly after oxidation and protonation, a constant decrease of the rectification ratio, together with an increase of the resistance were observed over time, most probably due to exposure to air and humidity. Unfortunately, a second protonation does not permit the recovery of the initial rectification ratio. This is likely to be due to the degradation of electrical contacts onto HNWs.

While asymmetric HNWs display rectifying characteristic in a doped state in the following we show that charge transport properties of as-synthesized HNWs are dominated by the bulk tridimensional hopping mechanism taking place in CP segments independently from their (a)symmetry and composition with up to five metal/CP segments.

Charge transport in CPs was described with a large number of theoretical models such as fluctuation-induced tunneling [48,49] space-charge limited current [38], Coulomb blockade [27,50], Charging energy limited tunneling [21] or by the critical regime near a metal-insulator transition [22,50]. Still, none of the above models explains satisfactorily the charge transport properties of as-synthesized HNWs. In a two- and four-probe configuration, they exhibit symmetric and superlinear *I-V* curves for *T* <=> 120 K. The *T*-dependence excludes an interface-related mechanism and is consistent with a smooth transition from an activated variable-range hopping (VRH) mechanism to an electric field ($F$) -induced hopping mechanism dominating in the low-$T$ regime ($T \leq 30$ K). Typical $I$-$V$ characteristics of as-synthesized tri-, tetra- and penta-segmented HNWs are given for different $T$ in Fig. 5. The $I$-$V$ curves are symmetrical and Ohmic between room temperature and $T_{NL} \approx 120$ K . Below $T_{NL}$, reversible nonlinearities appear and strengthen down to a low temperature regime ($T < 30$ K) where a zero current plateau appears around zero bias and the current becomes T-independent at higher bias (Fig. 6c). This effect could stem from a saturation of the sample resistance due to a self-heating effect. However, the same behavior is found in both vertical and planar devices featuring different thermal configurations. From the $I$-$V$ relationship, we deduce the resistance values $R(T)$ from the $(dI/dV)^{-1}$ values at zero bias. Despite the non-linearity, $R(T)$ is measured below $T_{NL}$ until the zero-current plateau appears. The corresponding $T$-dependence of the resistance are given in Fig. 5(a) for various tri-, tetra- and penta-segmented samples. Notably, the same behavior is observed for symmetric and asymmetric HNWs, independently from the number or composition of the segments. Similar $R(T)$ dependence with a drastic change of the $I$-$V$ relationship recorded at low-$T$ was observed in various CP systems [9,12,19,26,52]. To explicitly describe the behavior of $R(T)$ one can define a reduced activation energy $W(T) = -d(\ln R(T))/d(\ln T)$ [35]. In the insulating regime, $W(T)$ presents a negative temperature coefficient ; in the critical regime, $W(T)$ is nearly temperature independent while in the metallic regime, it has a positive temperature coefficient.

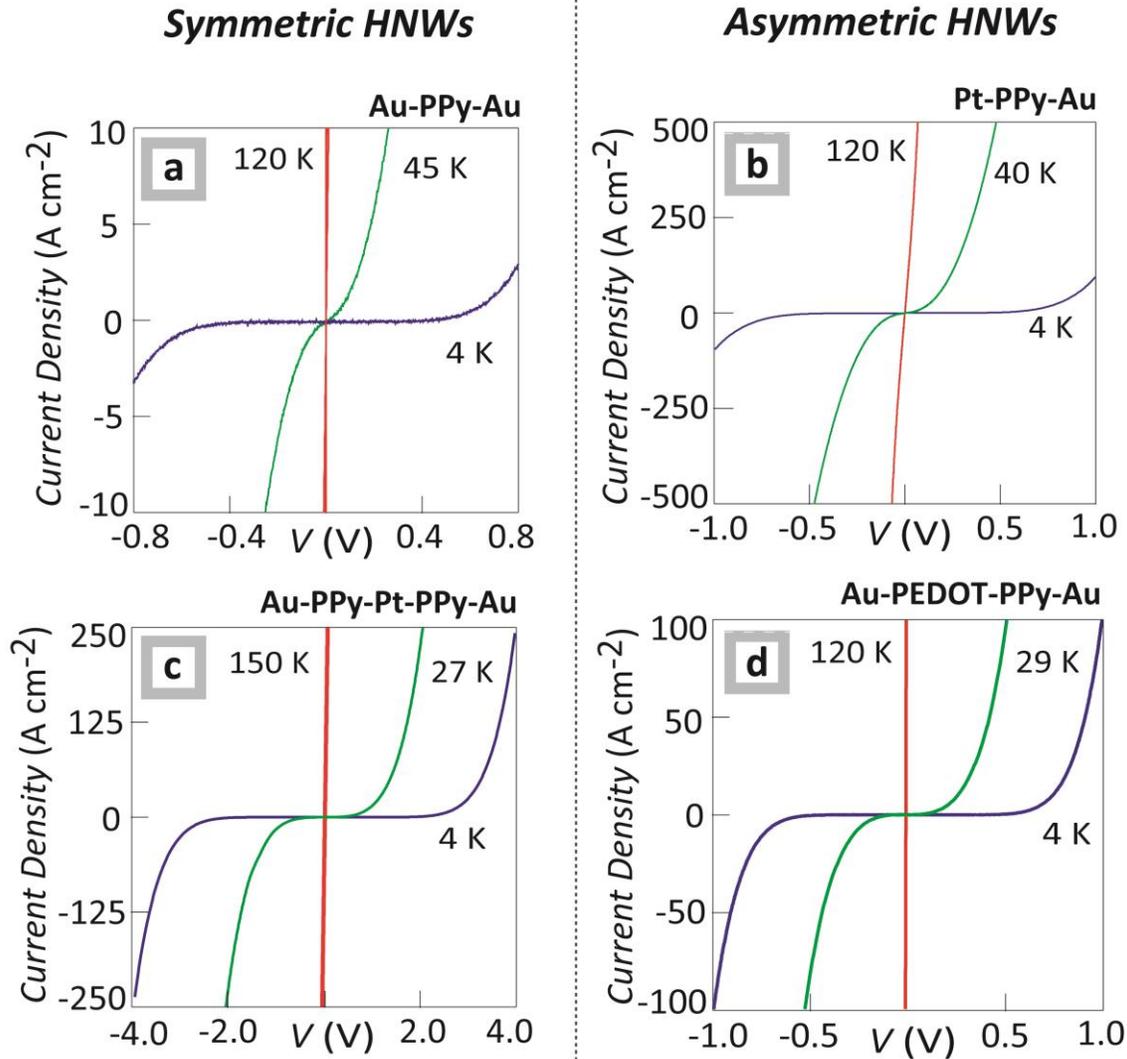

**Figure 5** *I-V curves of as-synthesized tri-, tetra- and penta-segmented HNWs. T-induced nonlinearities are symmetric and reversible for all HNW structures.*

The log-log plot of $W(T)$ is given in the inset of Fig. 6(a) for various HNW structures. It indicates that the conduction regime is clearly insulating whatever the HNW structure. Moreover, $R(T)$ follows the exponential T-dependence of the VRH model given by

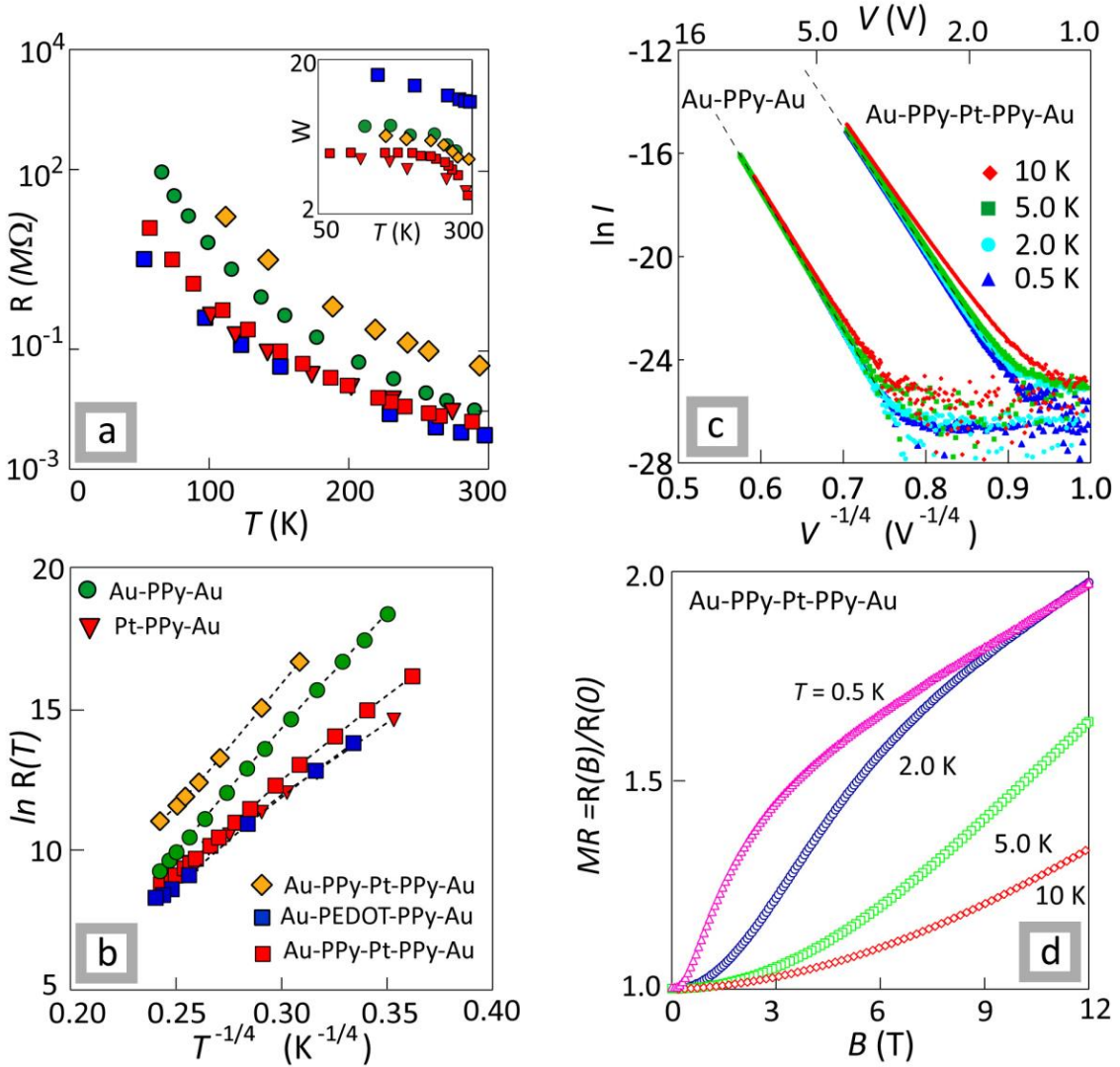

**Figure 6 (a) T-dependence of the zero-bias resistance of several HNW structures and (b) three-dimensional VRH plots. (c) I-V characteristics plotted as ln I versus $F^{-1/4}$ for two samples in the F-induced hopping regime $\beta \gg 1$ (see text). (d) Magnetoresistance of a penta-segmented HNW structure at 10, 5, 2 and 0.5 K. The dashed lines are fits to the data.**

$$\ln R(T) \propto (T_0/T)^{1/(d+1)} \qquad (2)$$

where $T_0 = 18\alpha^d/k_B N(E_F)$ is a characteristic temperature. The average hopping distance $R_{hop}$ is

$$R_{hop} = 1/2\alpha^{-1}(T_0/T)^{1/(d+1)} \qquad (3)$$

where $\alpha^{-1}$ is the localization length, $N(E)$ is the density of states at the Fermi energy, $k_B$ is the Boltzmann constant and $d$ is the dimensionality of the system. Here, $d$ and $T_0$ are determined from the best linear fit to Eq. 2. However, the determination of d is sometimes delicate because of the very narrow $T$ (typically one order of magnitude) range covered in most low-$T$ experiments [53] especially concerning organic materials [12,16,52]. To overcome this difficulty, we analyze two different hopping regimes observed for HNWs as a function of the temperature. In the high-$T$ range ($T_{NL} < T$) hopping is $T$-activated while in the low-$T$ range ($T \ll$

$T_{NL}$) the electric field ($F$)-induced hopping mainly contributes to the charge transport. To delineate the two hopping regimes, we define the parameter $\beta = |eR_{hop} F|/k_B T$. In the high-$T$ range, the energy gain at each hop due to $F$ is small compared to the thermal energy background which leads to $\beta \ll 1$. The field effect is thus negligible and the zero bias resistance $R(T)$ follows perfectly the exponential relation (Eq. 3) expected for the VRH mechanism, as shown in Fig. 5.(b) for various HNW structures. $\beta$ increases quickly when $T$ is lowered. In the low-$T$ range ($T < 30$ K), a zero-current plateau (figure 4) is observed for all HNW structures around the zero bias point because the thermal energy is too low to enable hopping. Therefore, charge transport solely occurs when $\beta \gg 1$ through a F-induced hopping mechanism. The current is then $T$-independent and follows the relation

$$\ln I \propto -(F_0/F)^{1/(d+1)} \qquad (4)$$

where $F_0 = k_B T_0/e\alpha$ stands for the critical electric field [54a,55]. As shown in Fig. 5.(c), a linear behavior is observed at the lowest temperatures where $\beta > 1$. $F_0$ are obtained from the slope of a linear fit to the data (values are listed in table 1). In the intermediate $T$ range, Ohmic $I$-$V$ relation is expected around the zero bias point while for increasing bias, superlinear $I$-$V$ curves should occur when $\beta > 1$. Remarkably, such behavior should be observed also at $RT$, however, the electric field to be applied to reach $\beta > 1$ at $RT$ requires an extremely high current density that irreparably damage samples. One should note that such $T$–independent $I$-$V$ curves are generally ascribed to self-heating phenomena. However the strong response of the magnetoresistance MR($B$) signal $vs$ $T$, shown in Fig. 6(d) and the good linear relationship of MR $vs$ $T^{1/4}$ demonstrate clearly that self-heating effect can be neglected in our measurements. From the analysis of the two hopping regimes we deduce $T_0$ and $F_0$. A third independent measurement is required to allow the retrieval of the parameters of the VRH model: $d$, $\alpha^{-1}$ and $N(E_F)$. We thus performed magnetoresistance experiments on the different HNW structures. A typical set of data is given in Fig.6(d) for a penta-segmented HNW. For every HNW structure, the measured MR is always symmetric, positive and it increases as $T$ is lowered. This behavior is well described by the wave-function shrinkage (WFS) model developed by Shklovskii and Efros [56] describing a parabolic-like MR for $B < B_c$ and a $B^{1/3}$ dependence at larger B with the critical magnetic field $B_c$ = 6h/[e$\alpha^2(T_0/T)^{1/4}$].

**Table 1. Data analysis for tri- tetra-, and penta-segmented samples. The Mott temperature $T_0$ was obtained by fitting the ln[$R(T)$] versus $T^{-1/4}$ plots and $F_0$ was extracted from linear fits to $ln\,I \propto (F_0/F)^{1/4}$ data (see text).**

| HNW As-synthesized ? Structures | $T_0$[K] | $F_0$ [V/μm] | $d$ | $\alpha^{-1}$ (nm) | $N(E_F)$ [eV$^{-1}$.cm$^{-3}$] | $R_{hop}$(300K) [nm] |
|---|---|---|---|---|---|---|
| **Au-PPy-Au** | $2.1\times10^8$ | $1.8\times10^7$ | 3 | 2.1 | $9.5\times10^{17}$ | 17 |
| **Pt-PPy-Au** | $7.6\times10^7$ | $4.9\times10^5$ | 3 | 2.4 | $2.0\times10^{17}$ | 26 |
| **Au-PEDOT-PPy-Au** | $1.5\times10^7$ | $1.6\times10^6$ | 3 | 1.6 | $3.0\times10^{18}$ | 12 |
| **Au-PPy-Au-PPy-Au** | $1.8\times10^8$ | $1.5\times10^7$ | 3 | 2.0 | $1.3\times10^{18}$ | 16 |
| **Au-PPy-Pt-PPy-Au** | $1.4\times10^7$ | $2.0\times10^6$ | 3 | 2.3 | $1.1\times10^{18}$ | 17 |

From $F_0$, $T_0$ and the $B_c(T)$ dependence, $d$, $\alpha^{-1}$ and $N(E_F)$ can be estimated. The obtained values are listed in table 1 for the HNW structures. It is worth noting that even taking into account the uncertainties on the parameters $F_0$, $T_0$ and $B_c$ $(T)$, the localization lengths deduced within the framework of the WFS model have very similar values and the small $\alpha^{-1}$ value reflects the strong disorder and low-doping level of the as-synthesized CP NWs. It is worth of note that the EBL processing of HNWs in a top-contact electrode configuration may have an impact on the doping level and disorder of CP segments due to electron beam exposure. However these effects were not considered here. Remarkably, the hopping mechanism is found for every HNW structure to be three-dimensional ($d$=3). This behavior is expected when charge carriers localized on a length much smaller than the NW dimensions [24]. This disorder is corroborated by the selected area electron diffraction patterns (SAED, not shown) obtained for each measured HNWs, which indicate an amorphous structure.

It can be shown that considering a narrow $T$-range, the $T$ and bias dependence of the HNW conductance, exhibit apparent power-law dependences similarly to what was found in a recent study for pure CP NWs [52]. However, considering broad $T$ and $F$ ranges, the stretched exponential dependences (Eq.2 and Eq.4) respectively observed in the Ohmic and non-Ohmic regimes ($\beta \gg 1$) better fit our set of data. As expected, in either regimes, the hopping distance $R_{hop}(T)$ is always much smaller than the CP segments between 2 nearby electrodes (2-4 µm) [53]. Based only on three parameters d, $N(E_F)$.and $\alpha^{-1}$ the VRH model provides an excellent description of the conduction behavior in symmetric and asymmetric HNWs over a wide range of temperature and electric field over three and four orders of magnitude, respectively.

In this paper, we reported on the electrical properties of muli-segmented HNWs. Remarkably, after doping, HNWs exhibit structure-dependent electrical properties with strong nonlinear $I$-$V$ characteristics. A rectifying behavior was observed only in asymmetric HNWs with rectification ratio as high as 3600 in PPy/PANi based HNWs. We discussed the origin of RT nonlinearities comparing the two- and four-probe measurements of single HNWs and proposed a simple model based on dual back-to-back Schottky diodes to account for the observations. On the other hand, identical charge transport mechanisms are observed in as-synthesized symmetric and asymmetric HNWs. In either case, HNWs exhibit symmetric non-linear $I$-$V$ characteristics for temperatures below $T_{NL} \approx 120$ K. These nonlinearities are reversible and strengthen at lower $T$. The analysis showed that charge transport in HNWs is bulk-limited and consistent with the 3D-VRH model over a wide range of temperature, magnetic and electric fields. This work suggests the potential use of HNWs in functional devices like nanoscale flexible resistive memory devices and underlines the efficiency of the template strategy for the synthesis of functional metal/CP nanostructures.

## 3. Methods

**Synthesis.** The growth of HNWs was performed in a one-compartment cell by using a classical three electrode electrochemical set up, with a Pt disk counter-electrode, an Ag/AgCl reference electrode, and a 300-nm-thick Au layer, coated on one side of the PC membrane, acting as working electrode. Metal and polymer stripes are consecutively deposited within the pores of nanoporous templates. 21-µm-thick polycarbonate track-etched membranes with a pore density of $10^9$ pores/cm$^2$ and pore diameter ranging from 90 to 110 nm, supplied by it4ip, were used as nanoporous templates. Metal segments are obtained by cathodic reduction of the metals ions, and CP segments by anodic oxidation of the corresponding monomer. All electrochemical synthesis was performed with a CHI660B Electrochemical Workstation (CH Inc). Prior to each electrodeposition, the bath was bubbled with nitrogen for a few minutes to allow nitrogen to diffuse into the pores of the template. After the synthesis of each material segment, we removed the used plating solution from the membrane and immersed the latter into de-ionized water for 30 minutes in order to allow the remaining plating solution diffusing outside the pores.

PPy segments were prepared by cycling voltammetry from an aqueous bath containing 0.02 M pyrrole and 0.1 M $LiClO_4$. The potential of the working electrode was swept between 0 and 0.85 V at a scan rate of 400 $mV.s^{-1}$. PEDOT segments were grown from an aqueous solution containing 14 mM of EDOT and 0.1 M $LiClO_4$, by sweeping the potential between 0.2 and 1.1 V at a scan rate of 400 $mV.s^{-1}$. The synthesis of PANI segments was performed using an aqueous solution containing 0.1 M aniline and 1 M HCl, by sweeping the potential from 0 to 0.9 V at a scan rate of 200 $mV.s^{-1}$. The electrodeposition of Au segment were carried out by cycling the potential from 0.7 to 0 V at 200 $mV.s^{-1}$ and using an aqueous bath containing 0.03 M $HAuCl_4.3H_2O$, 0.1 M KCl and 0.1 M $K_2HPO_4$. The growth of Pt segments was performed potentiostatically at -0.2 V using an aqueous solution containing 0.01 M $Na_2PtCl_6.H_2O$ and 0.5 M $H_2SO_4$. Pyrrole (Py, Acros, 99 %), 3,4-ethylenedioxyhiophene (EDOT, Sigma-Aldrich) and aniline (Acros, 99 %) were systematically purified through a microcolumn consisting of a Pasteur pipette, glass whool, and activated alumina before use. Lithium perchlorate ($LiClO_4$, Acros), hydrochloric acid (HCl, Acros), hydrogen tetrachloroaurate trihydrate ($HAuCl_4.3H_2O$, Acros, 99.9 %), sodium hexachloroplatinate ($Na_2PtCl_6$, Aldrich), potassium chloride (KCl, Acros), potassium hydrogenophosphate ($K_2HPO_4$), sulfuric acid ($H_2SO_4$, Fisher Scientific) and dichloromethane ($CH_2Cl_2$, Acros) were used without any prior purification. All aqueous solutions were prepared from de-ionized water. HNWs were released from the PC membrane by removing the evaporated Au layer with aqua regia or iodine solution, and dissolving the template in $CH_2Cl_2$. We have electrically characterized HNWs in both vertical (HNWs embedded in the PC membrane and single HNW devices) and planar configurations. In vertical configuration, measurements are performed on parallel HNWs (typically $10^4$ HNWs).

**Electrical Measurements.** In vertical configuration, a bunch of HNWs is electrically contacted to the pins of glass epoxy component mounting board using silver paint and 50 µm diameter gold wires immediately after the electrochemical synthesis. The NWs enclosed within the template are mechanically supported by the PC membrane and are protected from the effect of both oxygen and ambient humidity [55].

To enable the probing of HNW response to chemical doping, we also contacted single HNWs in planar configuration as follows. The PC membrane was dissolved and HNWs were dispersed in dichloromethane. Drops of this suspension were deposited onto home- made membrane-based Microdevices. After drying, the device was rinsed thoroughly with pure dichloromethane. HNWs remain attached to the substrate surface via van der Waals forces and their positions are then recorded by optical and/or electron microscopy. Au electrical probes are subsequently deposited by electron beam lithography (EBL) and lift-off for each HNW. We performed the chemical doping of HNWs by immersing the single NW device into a $FeCl_3$ 0.5 M or a HCl solution for 1 hour.

**Microdevices were** fabricated via Si micromachining techniques and state-of-the-art nanoscale electron beam lithography. Microdevices offer a unique tool for studying single nanowires by allowing the correlation between the structural properties - as contemplated, for instance, by transmission and scanning electron microscopies - and the electrical characterizations of the same HNW.

ACKNOWLEDGMENT

We thank Etienne Ferain and it4ip company for supplying polycarbonate membranes. S.D.C. thanks the F.R.S.-FNRS for her Senior Research Associate position.

76(11):115415, 2007.

in disordered systems, ii. Philo. Mag., 31:1327, 1975.